\documentclass[prd,twocolumn]{revtex4-1}
\usepackage{graphicx, epsfig}
\usepackage{color}
\usepackage{mathrsfs}
\usepackage{amsmath}
\usepackage{amsfonts}
\usepackage{bm}

\newcommand{\be}{\begin{equation}}
\newcommand{\ee}{\end{equation}}
\newcommand{\bea}{\begin{eqnarray}}
\newcommand{\eea}{\end{eqnarray}}
\newcommand{\ba}{\begin{eqnarray}}
\newcommand{\ea}{\end{eqnarray}}

\newcommand{\gapp}{\mathrel{\raise.3ex\hbox{$>$}\mkern-14mu
              \lower0.6ex\hbox{$\sim$}}}
\newcommand{\lapp}{\mathrel{\raise.3ex\hbox{$<$}\mkern-14mue
              \lower0.6ex\hbox{$\sim$}}}

\begin{document}
%\title{Magnetogenesis at Baryogenesis}
\title{Spectra of Magnetic Fields Injected during Baryogenesis}

\author{Yifung Ng$^1$ and Tanmay Vachaspati$^{1,2}$}
\affiliation{
$^1$CERCA, Department of Physics, 
Case Western Reserve University, Cleveland, OH~~44106-7079\\
$^2$Institute for Advanced Study, Princeton, NJ 08540\\ 
}

%%%%%%%%%%%%%%%%%%%%%%%%%%%%%%%%%%%%%%%%%%%%%%%%%%%%%%%
\begin{abstract}
\noindent
Helical magnetic fields are injected into the cosmic medium during 
cosmological baryogenesis and can potentially provide a useful probe 
of the early universe. We construct a model to study the injection 
process during a first order phase transition and to determine the 
power spectra of the injected magnetic field. By Monte Carlo 
simulations we evaluate the Fourier space symmetric and helical 
power spectra of the magnetic field at the time the phase transition
completes. The spectra are peaked at the scale given by the inverse 
size of bubbles at percolation and with a comparable width. These 
injected magnetic fields set the initial conditions for further 
cosmological magneto-hydrodynamical evolution.
\end{abstract}
%%%%%%%%%%%%%%%%%%%%%%%%%%%%%%%%%%%%%%%%%%%%%%%%%%
%\pacs{???}

\maketitle

Our study of the universe relies on relics left-over from early
cosmological epochs. The cosmic microwave background brings
information from the epoch when atoms formed, the light elemental
abundances from the epoch when nuclei formed. Similarly, the 
electroweak phase transition may mark the epoch when a net 
baryon number was generated, or when net lepton number
was converted into net baryon number, and this coincides 
with the generation of magnetic fields. The present paper 
is based on the hypothesis that primordial magnetic fields 
will inform us of the epoch when a net amount of baryons 
first formed, the so-called epoch of ``baryogenesis''
\cite{Cornwall:1997ms,Vachaspati:2001nb}.

The question of whether a primordial magnetic field exists is 
often raised in connection with the magnetic fields observed in 
galaxies and clusters of galaxies with strength $\sim \mu {\rm G}$
and kpc-Mpc coherence scale. There is considerable 
debate whether primordial fields are essential to the generation
of galactic fields, and what properties of the primordial
field are necessary to turn them into observed magnetic 
structures. The arguments involve 
the coherence and amplitude of observed magnetic fields, the 
efficiency of galactic dynamos, the turnover time scales 
associated with galactic dynamics, especially with the earliest 
known galaxies containing magnetic structures, astrophysical 
sources {\it e.g.} active galactic nuclei that may spew out 
magnetic fields, and the generation of large scale seed fields 
by the Biermann battery. We shall by-pass these issues since, 
in our view, a primordial magnetic field is of interest in itself, 
whether or not it is responsible for the observed magnetic fields 
in galaxies. If there are strongly motivated early universe 
scenarios, based on reasonably well-established particle physics,
that lead to the generation of magnetic fields, they provide good 
reason to study and to look for these structures in cosmological 
data. 

The connection between baryon number production and primordial
magnetic fields can be understood intuitively in the following 
way. Baryon number violation in the standard model of the 
electroweak interactions is made possible due to a quantum 
anomaly. As a physical process, baryon number is generated
when many different particles come together to form an object 
called a ``sphaleron'' \cite{Manton:1983nd,Klinkhamer:1984di}, 
which then decays into a final state with baryon number that is 
different from that of the initial state. The sphaleron, and 
its deformations, are made up of scalar and electroweak gauge 
fields and may also be viewed as being an unstable bound state 
of an electroweak magnetic monopole and an antimonopole, with
an electroweak $Z-$string confining them 
\cite{Vachaspati:1994ng,Hindmarsh:1993aw}. 
The monopole and antimonopole attract each other by the Coulomb 
force and the confining $Z-$string also pulls them together,
and a static solution is in general not possible. In the 
sphaleron though, the antimonopole has a twist in field space 
relative to the monopole which prevents the monopole and 
antimonopole from annihilating, and allows for the existence 
of a static solution \cite{Taubes:1982ie,Manton:1983nd}.
The presence of magnetic charges also explains the large
magnetic moment of the sphaleron calculated in 
Ref.~\cite{Klinkhamer:1984di}.
Twisted solutions similar to the sphaleron also occur in
the context of kinks in one spatial dimension and can lead to
novel static phases containing a lattice of kinks and
antikinks \cite{Vachaspati:2006zz}.

Once we appreciate that baryon number violation processes have 
intermediate states that consist of monopole-antimonopole 
pairs, it is not hard to see that magnetic fields must be
produced when a sphaleron decays. But there is a further
``twist'' to this connection. The twist in the fields that 
prevents the monopole and antimonopole from
annihilating is not stabilized, and the monopole can untwist 
and annihilate the antimonopole. The instability of the system 
causes the sphaleron to decay and radiate away its energy, 
releasing magnetic fields in the process 
\cite{Copi:2008he}. The decay has been studied
numerically and a very interesting feature emerges at late
times. The released magnetic fields inherit the twist of
the sphalerons and is measured by the magnetic helicity
integral
\begin{equation}
{\cal H} = \int d^3 x {\bf A}\cdot {\bf B}
\label{helicity}
\end{equation}
At late times, the magnetic field evolves such that the magnetic 
helicity is conserved. The conservation of magnetic helicity is 
familiar in magneto hydro-dynamics (MHD) in plasmas with high 
electrical conductivity. Yet in \cite{Copi:2008he} there is no 
external plasma; the only charges in the system are those resulting 
from the decay of the sphaleron itself. Remarkably, magnetic helicity 
is still conserved during sphaleron decay.

To summarize, baryon number violating processes in the electroweak
model occur via the production and subsequent decay of sphalerons.
Each sphaleron produces helical magnetic fields when it decays
and the helicity of the magnetic field is conserved. Now, since
the production of each baryon gives a certain amount of magnetic
helicity, the cosmological baryon number density, $n_b$, can be 
related to the magnetic helicity density ($h$) 
\begin{equation}
h \approx - n_b
\label{hnb}
\end{equation} 
where the minus sign requires more detailed considerations
\cite{Vachaspati:2001nb} and indicates that the primordial magnetic
field is left-handed. (In writing Eq.~(\ref{hnb}) we are assuming
that the cross-helicity between the magnetic field produced by
different sphalerons averages out to zero.) In principle, there 
could be a numerical pre-factor on the right-hand side and, in 
fact, early estimates suggested it to be $\sim \alpha^{-1}$ where 
$\alpha = 1/137$ is the fine structure constant. However, explicit 
numerical evolution of the sphaleron suggests that (\ref{hnb}) may 
be a better estimate \cite{Copi:2008he}. Numerically, the cosmic 
baryon number density is $\sim 10^{37}/{\rm cm}^3$ at the electroweak 
epoch.

The rough equality of magnetic helicity and baryon number gives
us an estimate for the integral in Eq.~(\ref{helicity}) but does
not provide much information about the characteristics of the
magnetic field itself. Of interest are the two point correlation
functions of the magnetic field. Assuming statistically 
homogeneous and isotropic magnetic fields, the spatial correlator 
at a fixed time can be written as
\begin{eqnarray}
C_{ij}({\bf r}) &\equiv&
\langle B_i ({\bf x}) B_j({\bf x} +{\bf r}) \rangle 
\nonumber \\
&=& \frac{1}{V} \int d^3x ~ B_i ({\bf x}) B_j({\bf x}+{\bf r})
\nonumber \\
&=& M_N(r) P_{ij} + M_L (r) {\hat r}_i{\hat r}_j  
+ \epsilon_{ijk}{\hat r}_k M_H(r)
\label{spacecorrelator}
\end{eqnarray}
where, $V$ is the integration volume, $i,j,k =1,2,3$, $r=|{\bf r}|$, 
${\hat r}={\bf r}/r$, $\epsilon_{ijk}$ is the Levi-Civita tensor, 
$P_{ij}$ is the traceless projection tensor,
\begin{equation}
P_{ij} = \delta_{ij}-{\hat r}_i {\hat r}_j
\label{rprojector}
\end{equation}
and $M_N$, $M_L$ and $M_H$ denote the ``normal'', ``longitudinal''
and ``helical'' correlation functions respectively. Also note that 
the helical term in Eq.~(\ref{spacecorrelator}) has a factor of 
${\hat r}_k$ and not ${\bf r}_k$ in it, as is sometimes used.
The correlation functions will also depend on time but here 
we are only interested in the {\it injected} field
at the end of the electroweak phase transition. So it is
to be understood that the correlation functions are all 
evaluated at $t=t_{ew}$ where $t_{ew}$ denotes the time at 
which the phase transition is complete.

The normal and longitudinal correlation functions are not 
independent since the magnetic field is divergenceless, 
${\bm \nabla}\cdot {\bf B}=0$, and the relation between $M_N$ 
and $M_L$ is,
\begin{equation}
\frac{dM_L}{dr} = \frac{2}{r}(M_N - M_L)
\label{dMLMN}
\end{equation}

It is more conventional to work in Fourier space, where the correlator
can be written as
\begin{eqnarray}
\langle b_i^* ({\bf k}) b_j ({\bf k}') \rangle &=& \nonumber \\
&& \hskip -1.5 cm
[S(k) p_{ij} + i \epsilon_{ijl}{\hat k}_l A(k)] ~
(2\pi)^3 \delta^{(3)}({\bf k}-{\bf k}')
\label{momcorrelator}
\end{eqnarray}
with no longitudinal term (proportional to ${\hat k}_i{\hat k}_j$) 
since the magnetic field is divergenceless.
We have denoted the Fourier components of ${\bf B}({\bf x})$ 
by ${\bf b}({\bf k})$ with the convention
\begin{equation}
b_i({\bf k}) = \int d^3x B_i({\bf x}) e^{+i{\bf k}\cdot {\bf x}}
\end{equation}
and the momentum space projection tensor is 
\begin{equation}
p_{ij} = \delta_{ij} - {\hat k}_i {\hat k}_j
\label{kprojector}
\end{equation}

By explicitly Fourier transforming $C_{ij}({\bf r})$ 
it is possible to derive
\begin{eqnarray}
S(k)&=& {4\pi} \int_0^\infty dr~ r^2 \biggl [
M_L \biggl ( \frac{\sin(kr)}{(kr)^3} - 
                \frac{\cos(kr)}{(kr)^2} \biggr ) 
\nonumber \\
&+&
M_N \biggl \{ \frac{\sin(kr)}{kr} - \frac{\sin(kr)}{(kr)^3}
              + \frac{\cos(kr)}{(kr)^2} \biggr \} 
\biggr ]
\label{SkfromMsfirst}
\end{eqnarray}
Further simplification occurs on using the divergenceless 
condition in Eq.~(\ref{dMLMN}), 
\begin{equation}
S(k)= {2\pi}\int_0^\infty dr~ r^2
\biggl ( \frac{\sin(kr)}{kr}-\cos(kr) \biggr ) M_L
\label{SkfromMs}
\end{equation}
where, to omit the boundary term, we have assumed $r^2 M_L(r) \to 0$ 
as $r \to \infty$. While Eq.~(\ref{SkfromMs}) is simpler, we still
use Eq.~(\ref{SkfromMsfirst}) in our numerical work because 
$r^2M_L(r)$ is not negligible at the scale of our simulation box.
Similarly we have the relation
\begin{equation}
A(k) = \frac{4\pi}{k} \int_0^\infty dr~ r ~ \left (
\frac{\sin(kr)}{kr} - \cos(kr) \right ) M_H 
\label{AkfromMH}
\end{equation}

The goal of this paper is to develop a model for the generation of 
magnetic fields during baryogenesis (Sec.~\ref{model}), to Monte Carlo 
the model (Sec.~\ref{montecarlo}) and to obtain the correlation
functions of the magnetic field (Sec.~\ref{results}). Our final 
results will be the functions $M_N(r)$, $M_L(r)$, $M_H(r)$, 
$S(k)$ and $A(k)$. The spectral properties of the injected field
should be useful to study their evolution. 
%Ultimately we would
%like to study the feasibility of observing baryogenesis generated 
%magnetic fields and also to consider magnetic fields as probes of
%the cosmological electroweak epoch. 
We discuss limitations of our 
model and future prospects in Sec.~\ref{conclusions}.

\section{Model}
\label{model}

The relation between baryogenesis and magnetic fields exists as long 
as there is anomolous baryon number violation. Even in leptogenesis
scenarios, sphaleron processes are required to convert lepton number 
into baryon number \cite{Davidson:2008bu}, and these conversions 
will produce magnetic fields. The power spectra of fields produced 
during leptogenesis, however, may be different from those produced 
during baryogenesis as considered here because, unlike baryogenesis, 
leptogenesis does not rely on a first order phase transition for
departures from thermal equilibrium. 

We assume that baryogenesis occurred at the electroweak scale
and the necessary departures from thermal equilibrium are provided
by a phase transition that was strong enough. The
electroweak symmetry is broken within bubbles which then
grow, merge and eventually fill all of space, thus completing the phase 
transition. During this process, baryon number changes can occur 
relatively freely outside the bubbles since there is little energy
cost associated with the sphaleron in the symmetric phase, but 
baryon number changes are highly suppressed within the bubbles since
the sphaleron has mass $\sim m_W/\alpha$ in the symmetry broken
phase. (Here $m_W \sim 100$ GeV is the mass of the electroweak W 
gauge boson.) Also, in the symmetric phase that is outside the 
bubbles, the electromagnetic magnetic field is just a linear 
combination of the electroweak magnetic fields, and has no special 
significance. For example, all the electoweak gauge fields are massless, 
and we expect rapid interactions to maintain equilibrium between the 
different degrees of freedom.  Inside the bubbles, the electromagnetic 
field is the only massless gauge field and its evolution is described 
by Maxwell's equations. However, there are no sphaleron transitions 
inside the bubble and so no magnetic fields are generated there. It 
is only the sphaleron transitions occuring in a thin region right 
around the bubble wall which produce the magnetic fields that are then 
captured by the growing bubble. It is this magnetic field, generated 
by sphaleron transitions at the surface of bubbles, that is of 
interest to us.

\begin{figure}
\scalebox{1.0}{\includegraphics[angle=0]{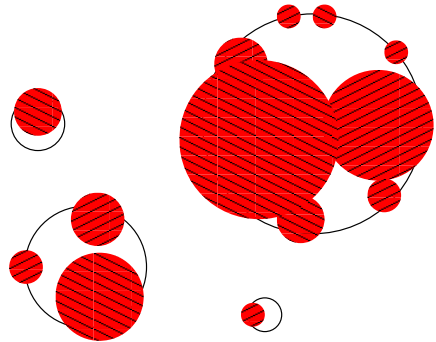}}
\caption{Schematic representation of the electroweak first order
phase transition via growing bubbles (circles), and of the sphaleron
explosion-like events that occur on the surfaces of the bubbles.
The discs represent the magnetic fields from the explosion event,
with earlier explosions having had more time to grow out further.
}
\label{schematics}
\end{figure}

A schematic picture of our model is shown in Fig.~\ref{schematics}.
Essentially, there are bubbles of the broken symmetry phase that 
nucleate and grow, and sphalerons are explosions that occur on the 
surfaces of these bubbles and blow out magnetic fields into the 
environment. We would like to find the correlation functions of 
the magnetic fields left-over after the phase transition has 
completed. Our picture of the phase transition is similar
to that of cosmological large-scale structure formation in
which cosmic voids nucleate and grow. Supernovae or other 
astrophysical activity occurs on the surfaces and intersections 
of the voids and expels magnetic fields and other elements into 
the cosmic environment.

\begin{figure}
\scalebox{0.65}{\includegraphics[angle=0]{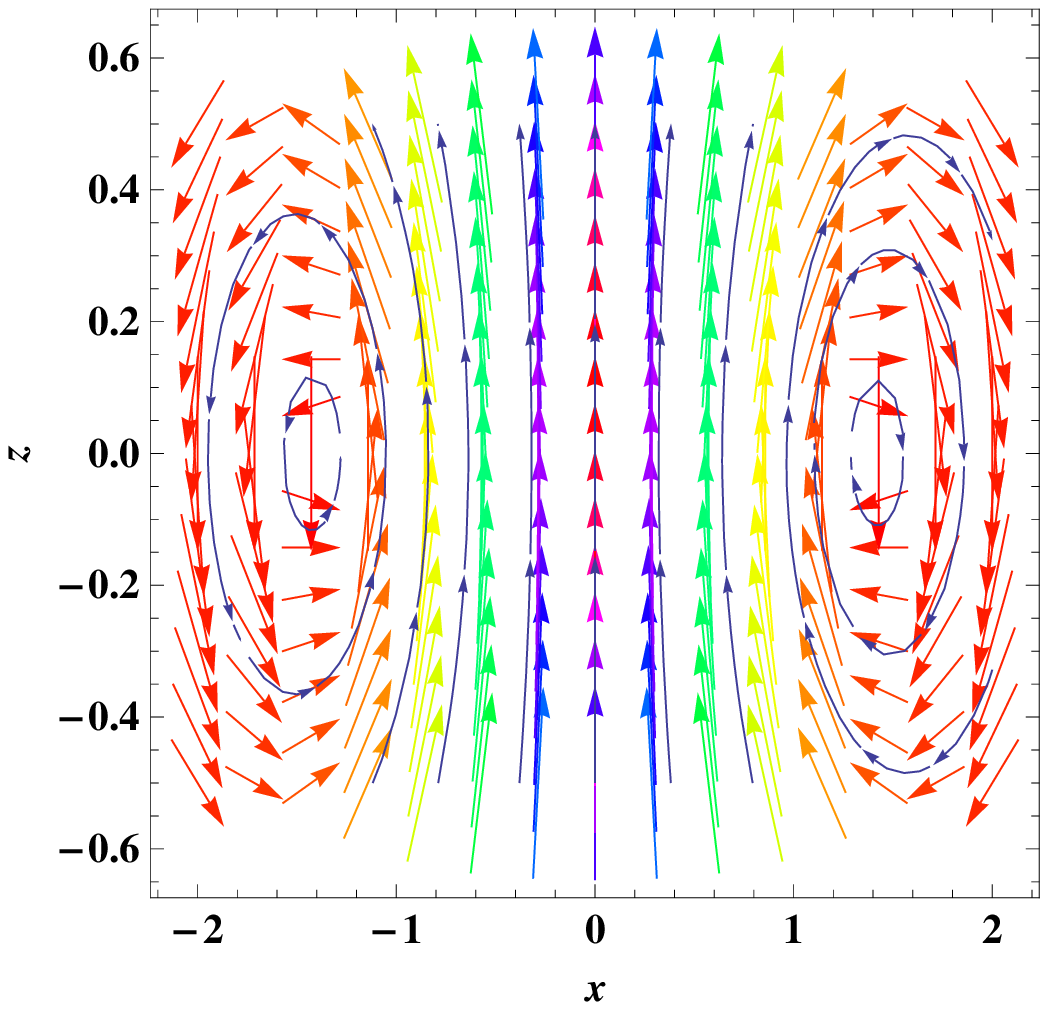}}
\scalebox{0.65}{\includegraphics[angle=0]{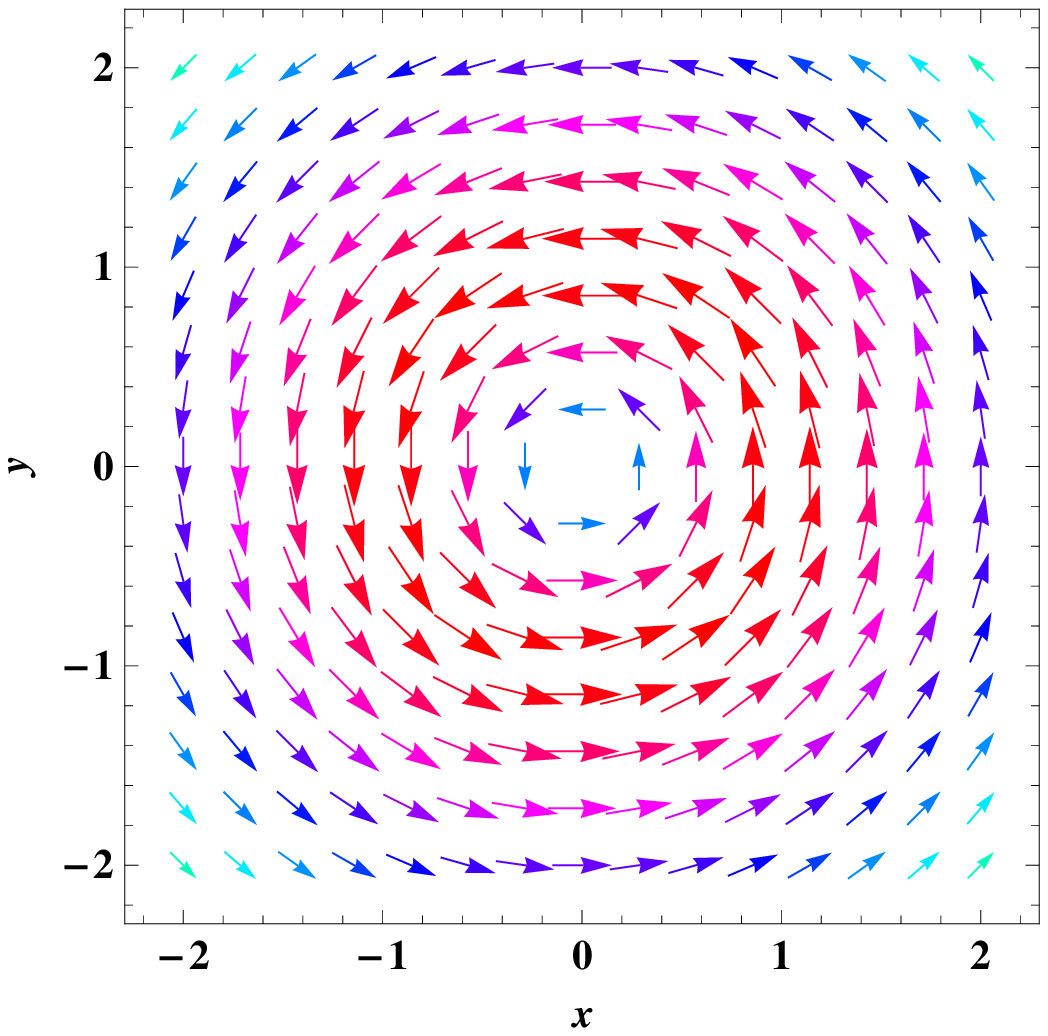}}
\caption{Plots of the magnetic field vectors in the xz-plane 
(top) and xy-plane (bottom) with $a=1$ in Eq.~(\ref{Bchoice}).
}
\label{B-plots}
\end{figure}

The only element that is missing from the picture so far is a model
for the ``explosion'' that expels the magnetic field. 
In Ref.~\cite{Copi:2008he}, the decay of the sphaleron was studied
numerically, with the result that magnetic helicity stayed constant
at late times, and the energy density spread out as time progressed. 
We shall model the magnetic field in a sphaleron explosion as
\begin{eqnarray}
B_r &=&  \frac{a \cos\theta}{(a^2+r^2)^{3/2}} \nonumber \\
B_\theta &=& - \frac{a \sin\theta}{2}
            \frac{2a^2-r^2}{(a^2+r^2)^{5/2}} 
\label{Bchoice}                                         \\
B_\phi &=& \frac{r}{a^3} e^{-r/a} \sin\theta \nonumber 
\end{eqnarray}
where $(r,\theta,\phi)$ are spherical coordinates centered at the
location of the sphaleron and the z-axis is chosen so that there 
is azimuthal symmetry. (Plots of two sections of the magnetic
field are shown in Fig.~\ref{B-plots}.) The $r,\theta$ components 
of the magnetic field are chosen to be approximately those of a 
current-carrying circular loop of wire of radius $a$, and with
current proportional to $1/a$.
(The approximation in Eqs.~(\ref{Bchoice}) for the field of a 
circular loop of wire is taken from Sec.~5.5 of Ref.~\cite{Jacksonbook}.)
The azimuthal component of the magnetic field is chosen so that 
it is localized in a region of size $\sim a$. The size of the
whole system is chosen to grow with time to model the exploding
sphaleron. Assuming that the magnetic fields generated 
by the sphaleron expand out at the speed of light, we can take 
$a(t)=t-t_0$ where $t_0$ is the epoch at which the sphaleron 
transition occurred.

%\begin{figure}
%\centering
%\begin{tabular}{cc}
%\epsfig{file=loop.eps,width=0.5\linewidth,height=0.5\linewidth} &
%\epsfig{file=torus.eps,width=0.5\linewidth,height=0.5\linewidth} \\
%\end{tabular}
%\caption{Schematic figures of the sphaleron model: $B^{(1)}$ and $B^{(2)}$}
%\end{figure}
%\vspace{2pc}

The scalings with $a$ are important since the magnetic helicity 
in the aftermath of a sphaleron decay stays constant, as we now 
show explicitly. We first construct the gauge potential for the 
azimuthal component of the magnetic field
\begin{equation}
A^{(\rm azim)}_\theta = 
- \frac{e^{-r/a}}{r} \left ( 2 + 2\frac{r}{a}
    + \biggl ( \frac{r}{a} \biggr )^2 \right ) \sin \theta
\label{Aazim}
\end{equation}
In other words, 
\begin{equation}
{\bm \nabla}\times (A^{(\rm azim )}_\theta {\hat \theta}) = 
B_\phi {\hat \phi}
\end{equation}
Now we define
\begin{equation}
{\bf B} = {\bf B}^{(1)} + {\bf B}^{(2)}
\end{equation}
where,
\begin{equation}
{\bf B}^{(1)} = B_r {\hat r} + B_\theta {\hat \theta}
\end{equation}
\begin{equation}
{\bf B}^{(2)} = B_\phi {\hat \phi} 
\end{equation}
and, correspondingly,    
\begin{equation}
{\bf A} = {\bf A}^{(1)} + {\bf A}^{(2)}
\end{equation}
so that ${\bf B}^{(i)} = {\bm \nabla}\times {\bf A}^{(i)}$.
Then since the helicities in ${\bf B}^{(1)}$ and ${\bf B}^{(2)}$ are 
individually zero, the total helicity is given by the cross terms
\begin{equation}
{\cal H} = \int d^3 x {\bf A}^{(1)} \cdot {\bf B}^{(2)}
          +\int d^3 x {\bf A}^{(2)} \cdot {\bf B}^{(1)}
\end{equation}
An integration by parts is now used to obtain
\begin{equation}
{\cal H} = 2 \int d^3 x {\bf A}^{(2)} \cdot {\bf B}^{(1)}
\end{equation}
The boundary term vanishes because the fields are localized.

Next we insert the expressions for the gauge potential and
magnetic field as given in Eqs.~(\ref{Bchoice}), (\ref{Aazim}). 
A simple change of variables,
$u=r/a$, shows that the magnetic helicity of one source is 
\begin{equation}
{\cal H} = \frac{8\pi}{3} 0.57 = 4.8 
\end{equation}
for any value of $a$. Therefore the magnetic helicity is
conserved for our choice of magnetic fields and matches
the conservation seen in sphaleron decay \cite{Copi:2008he}.

The choice of factors of $a$ in Eq.~(\ref{Bchoice}) also ensures 
that the relative energy in each of the three components of
the magnetic field stays fixed, while the net energy in the 
magnetic field decays as $1/a$.
The decay of the magnetic energy is a necessary consequence of
the conservation of helicity because the energy density is
${\bf B}^2$ while the helicity density is ${\bf A}\cdot {\bf B}$
and hence, simply by counting dimensions, the total energy is
the total helicity divided by a length scale. Since the only
length scale in the problem is $a$ and the total helicity
remains constant, the total energy must decay as $1/a$.

The magnetic field in Eq.~(\ref{Bchoice}) is axially symmetric
about the chosen z-axis. However, different sphalerons will
produce magnetic fields that are azimuthally symmetric with 
respect to different axes. So a sphaleron is described by its 
location as well as its orientation. We will assume that the
orientations of the sphalerons are isotropically distributed 
and, for a given sphaleron, choose the orientation from
a uniform distribution on the two-sphere {\it i.e.} in
spherical coordinates, $\cos\theta$ and $\phi$ are chosen
from a uniform distribution over the intervals $(-1,1)$
and $(0,2\pi)$ respectively. In principle, the interaction
of electroweak fields with the bubble wall could result
in a preferential orientation of the sphaleron (e.g. normal
to the bubble wall) but we shall disregard this possibility 
in the present paper.

We do not claim that an electroweak sphaleron produces the 
magnetic field in Eq.~(\ref{Bchoice}) when it decays. Instead, 
(\ref{Bchoice}) is a convenient choice for the magnetic field 
and has the following desirable properties: 
(i) The field is smooth and divergenceless.
(ii) The magnetic helicity is independent of $a$. 
(iii) The relative energy in all three components of the
magnetic field is independent of $a$. 
Furthermore, we are only interested in the correlation functions
for the magnetic field at large separations and hope that these
are not sensitive to the exact form of the model we choose for
the sphaleron's magnetic field. To put this work on a firmer 
footing, it will be necessary to study the process of sphaleron 
decay more carefully and to devise a more accurate model for 
the magnetic fields produced. 

\section{Monte Carlo Simulation}
\label{montecarlo}

The procedure we follow is to throw bubble sites randomly with
uniform distribution within our simulation volume. Successful
bubble sites are those that lie outside of all existing bubbles.
Then the bubbles grow at speed $v_b$. The bubble growth
velocity depends on the ambient plasma and can be much
smaller than the speed of light (e.g. \cite{Cline:2006ts},
\cite{Bodeker:2009qy}, \cite{Espinosa:2010hh}).
However, to keep the number of parameters to a minimum 
in our simulations, we took $v_b =c =1$. 
As the bubbles grow, we randomly nucleate sphalerons on the 
bubble surfaces with angular density $1/A_s$ where $A_s$ 
denotes the average area per sphaleron. Here also we take
care to eliminate sphaleron sites that lie within pre-existing
bubbles. Each sphaleron is 
described by its time of nucleation, location, as well as 
its orientation. The sphaleron nucleation time enters
the factor $a$ in Eq.~(\ref{Bchoice}), while the spatial
location sets the origin for the axes, and the orientation of
the sphaleron fixes the direction of the local $z$-axis. With
time, the radius of the magnetic field grows with velocity
$v_m$ which, for convenience, is also taken to be the speed
of light, $v_m=c =1$. After the phase transition is complete
-- no more bubbles nucleate since almost all the simulation
volume is occupied by pre-existing bubbles -- we find the magnetic 
field on a lattice within our simulation volume due to all 
sphalerons in our simulation box. We then compute the correlation 
function, $C_{ij} (r {\hat r})$ (Eq.~(\ref{spacecorrelator}),
which has $3\times 3\times 3$ components because of the 2 free 
indices $i,j$ and the 3 choices for the direction of ${\hat r}$.
The simulation is run many times with different 
seeds for the random number generator and ensemble averages are 
calculated. We have explicitly checked that the correlators are 
of the form in Eq.~(\ref{spacecorrelator}). Then the spectral 
functions are found as linear combinations of the 
$C_{ij}(r {\hat r})$,
\begin{eqnarray}
M_L(r)&=&\frac{1}{3} \sum_{\hat r} {\hat r}_i{\hat r}_j 
               C_{ij}(r{\hat r})\\
 M_N(r)&=&\frac{1}{6} \sum_{\hat r} P_{ij}  C_{ij}(r{\hat r})\\
 M_H(r)&=&\frac{1}{6} \sum_{\hat r} \epsilon_{ijk} {\hat r}_k
               C_{ij}(r{\hat r})
\label{projections}
\end{eqnarray}
where the sum is over the 3 directions: 
${\hat r}\in ({\hat x},{\hat y},{\hat z})$.

The simulation takes care that bubbles can only nucleate in
the false vacuum region, that is, outside every other bubble.
In practice, a certain number, $n_b$,  of bubbles are thrown down 
at every time step, and any bubbles that lie within pre-existing
bubbles are rejected. Next we want to locate sphaleron events 
on the surfaces of existing bubbles. We let $A_s$ denote the 
mean area occupied by a sphaleron on the surface of the bubble. 
On the surface of a bubble of radius $R$, we throw $4 \pi R^2/A_s$ 
sphaleron sites where we take $A_s =32$ so that there are a large 
number of sphalerons (order $10^5$) in the simulation but
still within computational limits. We reject those sites that lie 
within any other bubble. For large bubbles, the mean distance 
between neighboring sphalerons on the same bubble is 
$\approx 2\sqrt{A_s/(4 \pi)} \approx  3$. 

\begin{figure}
\scalebox{0.33}{\includegraphics[angle=-90]{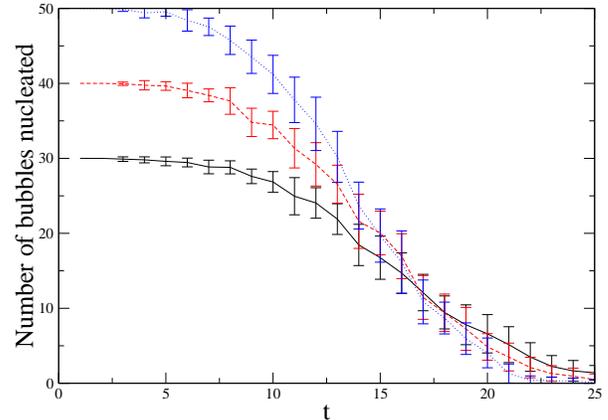}}
\caption{The number of bubbles nucleated at time $t$ for 3 
different nucleation rates. The nucleation 
rates correspond to throwing down 30 (black solid curve), 
40 (dashed red) and 50 (dotted blue)
bubble sites per time step. The bubbles rapidly fill 
the simulation box and even with the lowest nucleation rate 
the phase transition is essentially complete by $t=25$ 
{\it i.e.} 25 time steps. 
}
\label{nbubvst}
\end{figure}

\begin{figure}
\scalebox{0.33}{\includegraphics[angle=-90]{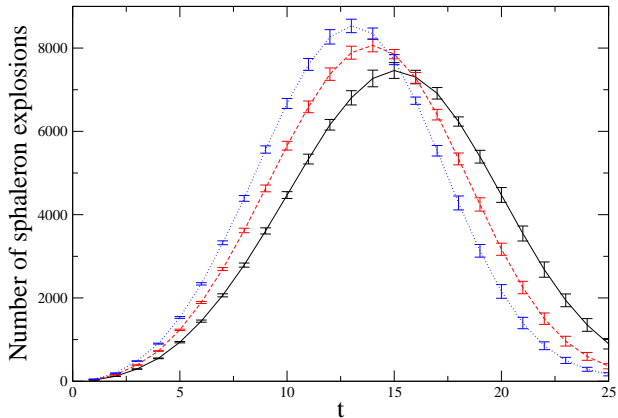}}
\caption{
The number of sphalerons nucleated at a time $t$ for the three 
different bubble nucleation rates as in Fig.~\ref{nbubvst}. 
}
\label{nsphvst}
\end{figure}

The bubble nucleation rate is chosen over a range that gives
$\sim 10^2$ bubbles in the simulation. The sphaleron nucleation 
rate is kept fixed, while the bubble nucleation rate is taken
to be 30, 40 and 50 bubbles per time step. The number of bubbles 
nucleated per time step for the three different nucleation rates 
are shown in Fig.~\ref{nbubvst}. The error bars denote 1$\sigma$ 
fluctuations about the mean taken over 20 runs. In Fig.~\ref{nsphvst} 
we plot the number of sphalerons nucleated per time step, also 
including 1$\sigma$ error bars. This plot is equivalent to plotting 
the surface area separating the true vacuum and the false vacuum
as a function of time.

A subtle point about the simulation is that we have nucleated
bubbles within a box but then the bubbles subsequently grow 
beyond the box. The sphalerons are, however, nucleated only within 
the box at a fixed rate, rejecting those sphalerons that lie on 
the parts of the bubbles that are outside the box. So the magnetic 
field close to the boundaries of the box suffer from boundary 
artifacts due to the lack of sphalerons outside the box. 
Hence it is important that the magnetic field only be calculated 
in a sub-box that is smaller than the original box, at least by 
a margin that is larger than the size of the typical magnetic 
structure, given by $a(t)$. In our simulations, the box size was 
144 lattice spacings and the sub-box size was 108 spacings, 
{\it i.e.} we excluded a boundary layer of 18 lattice spacings 
all around the box. As can be seen from Fig.~\ref{nsphvst}, most 
sphalerons nucleated at $t \sim 15$ and had $a \sim 10$ when we 
stopped the simulation ($t=25$). So $a$ for most bubbles is 
less than the thickness of the excluded boundary layer. 
The other relevant simulation parameters are the lattice 
spacing, $dx=1$, and the time step, $dt=1$.

After the phase transition is complete, we know where all the
sphalerons are located, their orientations, and also their
sizes because we know the times at which the sphalerons 
exploded. We take the expansion speed of the magnetic
fields produced by a sphaleron to be the speed of light.
This determines the size, $a(t)$, occurring in Eq.~(\ref{Bchoice})
for every sphaleron. Then at every point on a sub-lattice
we sum over the magnetic field due to every sphaleron.
(This is the computationally expensive part of the code
since it involves roughly $10^5\times (108)^3 \sim 10^{11}$
computations.) Once we know the magnetic field at each lattice
site, we calculate spatial correlations by doing the volume
integral in Eq.~(\ref{spacecorrelator}) and averaging over 20
ensembles. Projections of the correlation functions as in 
Eq.~(\ref{projections}) immediately give the normal,
longitudinal and helical power spectra. The integrals in 
Eq.~(\ref{SkfromMsfirst}), (\ref{AkfromMH}) finally lead to the 
Fourier space power spectra, $S(k)$ and $A(k)$.

%
%\begin{table}
%\center
%\begin{tabular}{|c|c|}
%\hline
%Physical Quantity @ EW & Numerical Value \\
%\hline
%$L_{\mathrm{EW}}$ & $10^{-18}m$ \\
%$T_{\mathrm{EW}}$ & $100GeV$ \\
%$t_{\mathrm{EW}}$ & $10^{-12}s~(T\approx \frac{MeV}{\sqrt{t}}$\cite{Kolb90}) \\ 
%$Horizon(\mathrm{t_{EW}})$ & $10^{-12}*10^{8}m=10^{-4}m$  \\
%$L_{\mathrm{Debye}}$ & $O(e^{2}T)=10^{-14}m$  \\
%$L_{\mathrm{sphaleron}}$ & $M_{W}^{-1} \approx 10^{-18}$m \\
%$h=\frac{1}{V}(\int{B_H\cdot A_{\mathrm{dipole}}dV})$ & $-\frac{n_B}{\alpha}=10^{44}m^{-3}$\cite{Copi:2008he}  \\
%$\frac{n_{b,T}}{s_{T}}$($s$=$\frac{2\pi^2g_*T^3}{45}$) & $10^{-10}$\cite{Kolb90} \\
%$n_{b,T_{EW}}$ & $10^{-40}{\rm cm}^{3}$\\
%\hline
%\end{tabular}
%\caption{Length scales of interest in electroweak baryogenesis model}
%\label{table:length}
%\end{table}

Before proceeding to the numerical details and results, we summarize 
the electroweak and cosmological parameters. The Hubble distance at 
the electroweak epoch is $H_{\rm ew}^{-1} \approx 10 ~ {\rm cm}$,
while the thermal length scale is 
$T_{\rm ew}^{-1} \approx 10^{-16} ~ {\rm cm}$. 
The sphalerons are exploding on the inverse electroweak
mass length scale which is comparable to the thermal length
scale. The ejected magnetic fields produced can spread out freely 
until the MHD frozen-in length scale at the electroweak epoch, 
$l_{\rm frozen} \approx \sqrt{t_{\rm ew}/4\pi \sigma_c}
\approx 10^{-8} ~ {\rm cm}$, where $\sigma_c \sim T_{\rm ew}/e^2$
is the electrical conductivity of the plasma. The present baryon 
number density is $n_b \sim 10^{-7} ~ {\rm cm}^{-3}$ and at the 
electroweak epoch this corresponds to 
$n_{\rm b,ew} \sim 10^{37} ~{\rm cm}^{-3}$.

\section{Results}
\label{results}

In Fig.~\ref{nsphvst} we have already shown the nucleation of 
sphalerons as a function of time for each of the different bubble
nucleation rates. (We do not vary the sphaleron nucleation rate.) 
The number of sphalerons nucleated at any time is proportional 
to the net surface area of the bubbles. As expected, the plot 
shows that the surface area, and hence the sphaleron nucleation
rate, grows to a maximum and then decreases. If we 
increase the bubble nucleation rate, the rate of growth of the 
surface area is larger initially, but then the turning point is 
at earlier times because it is determined by the merging of 
bubbles. 

\begin{figure}
\scalebox{0.33}{\includegraphics[angle=-90]{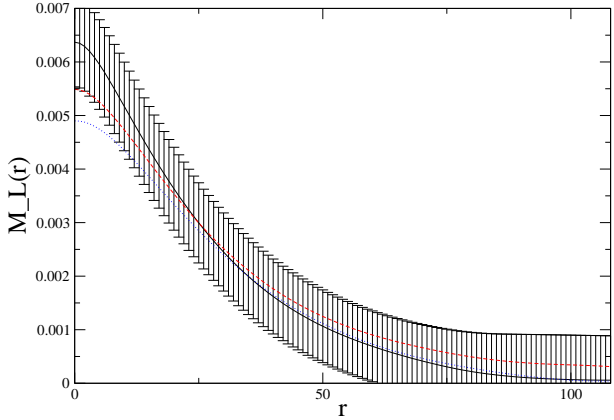}}
\caption{$M_L (r)$ for different rate 
of bubble nucleation.
Solid (black), dashed (red), and dotted (blue) 
curves correspond to 30, 40 and 50 
bubbles nucleated at every time step. Fluctuations are only 
shown for the run with 30 bubbles nucleated per time step.
}
\label{MLvsr}
\end{figure}

\begin{figure}
\scalebox{0.33}{\includegraphics[angle=-90]{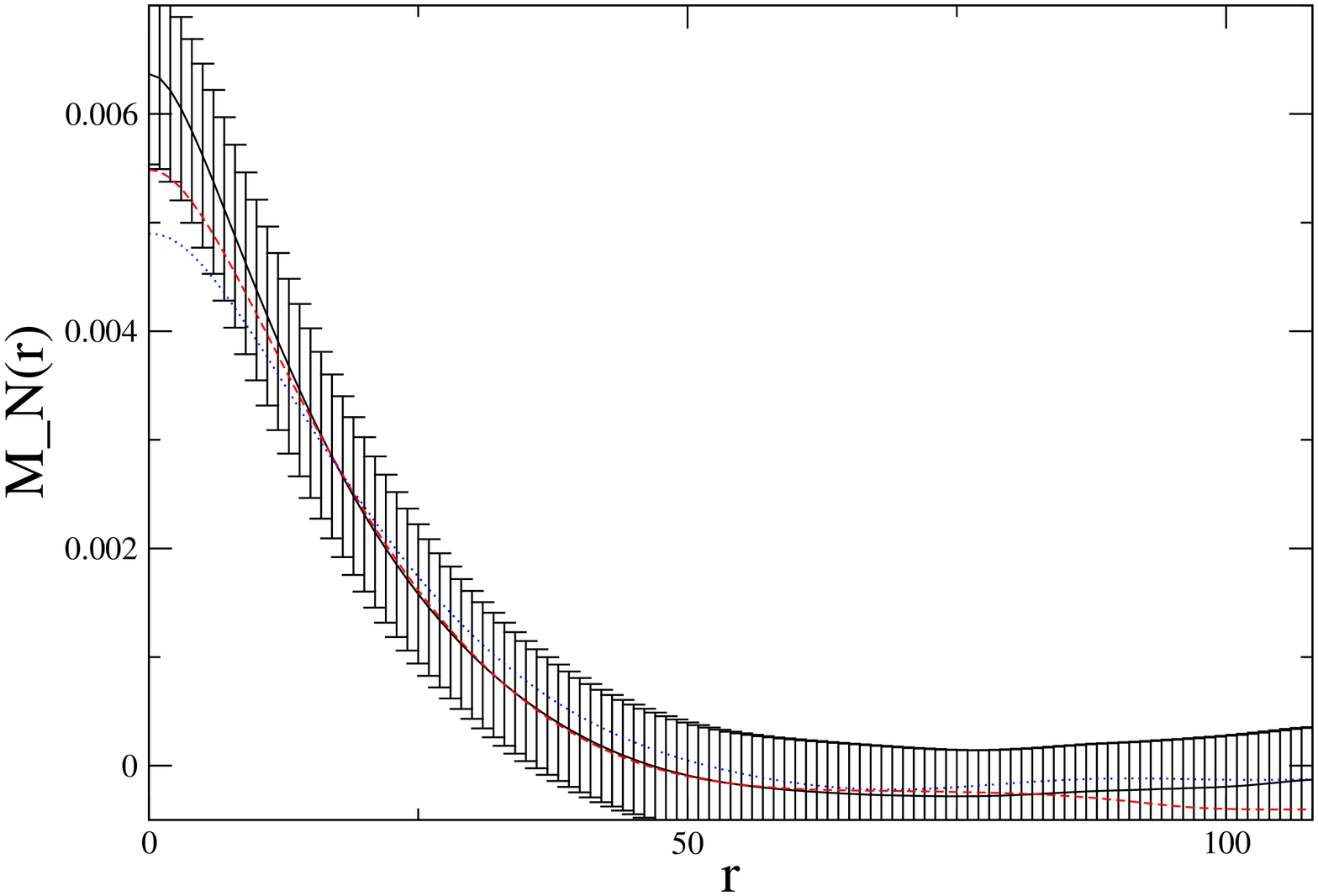}}
\caption{$M_N(r)$. 
Plots are made following the scheme of Fig.~\ref{MLvsr}.
}
\label{MNvsr}
\end{figure}

\begin{figure}
%\centerline{\scalebox{0.8}{\input{feynmandiag.pstex_t}}}
\scalebox{0.33}{\includegraphics[angle=-90]{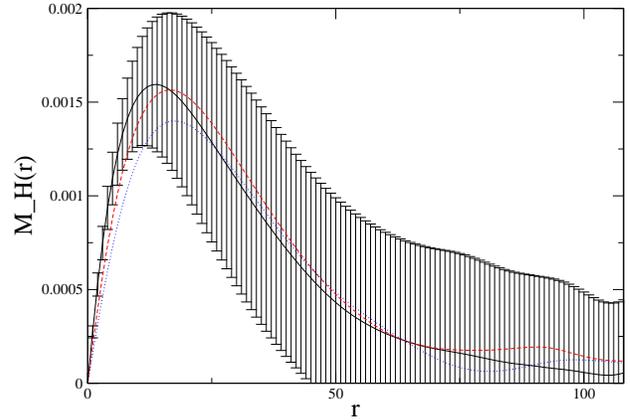}}
\caption{$M_H(r)$.
Plots are made following the scheme of Fig.~\ref{MLvsr}.
}
\label{MHvsr}
\end{figure}

The plot of the sphaleron rate versus time in 
Fig.~\ref{nsphvst} also tells us the size distribution
of the expanding magnetic field distribution. In particular,
lower bubble nucleation rates, as are relevant in strongly first
order phase transitions, lead to a sphaleron rate that is larger
at later times. (The peak in Fig.~\ref{nsphvst} is shifted 
to the right.) So at some fixed late time, the sphaleron 
explosions have had less time to grow and there are a larger 
number of smaller sphaleron explosion remnants. The correlation
functions should therefore be larger at small distances when
the bubble nucleation rate is smaller. This feature can be
seen in Figs.~\ref{MLvsr} and \ref{MNvsr} where we show 
the spatial correlation functions, $M_L$ and $M_N$, 
versus $r$. In Fig.~\ref{MHvsr} we show the helical 
correlation function, $M_H(r)$. The peak shifts to the
left for smaller bubble nucleation rates in agreement
with our observation above that more sphaleron explosions
occur later if the bubble nucleation rate is small
and have less time to grow. 

The fluctuations in the correlation functions, denoted by the 
error bars, are quite large. To further reduce them would require 
increasing the sphaleron rate and would increase the computational 
time. At present, each Monte Carlo with 20 runs takes about
10 days to run.

\begin{figure}
\scalebox{0.33}{\includegraphics[angle=-90]{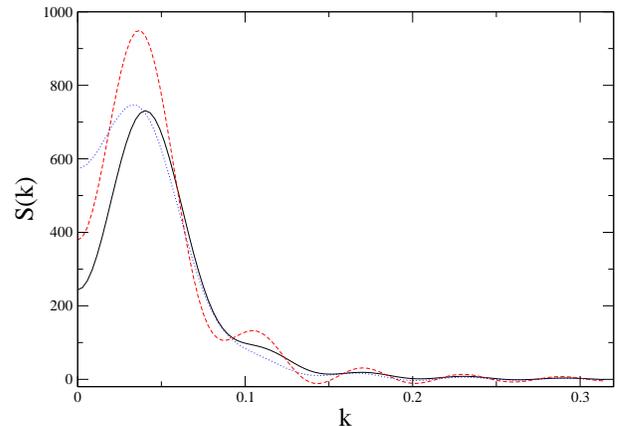}}
\caption{$S(k)$ vs. $k$. Plots are made following the scheme of 
Fig.~\ref{MLvsr}.
}
\label{Sofk}
\end{figure}

\begin{figure}
\scalebox{0.33}{\includegraphics[angle=-90]{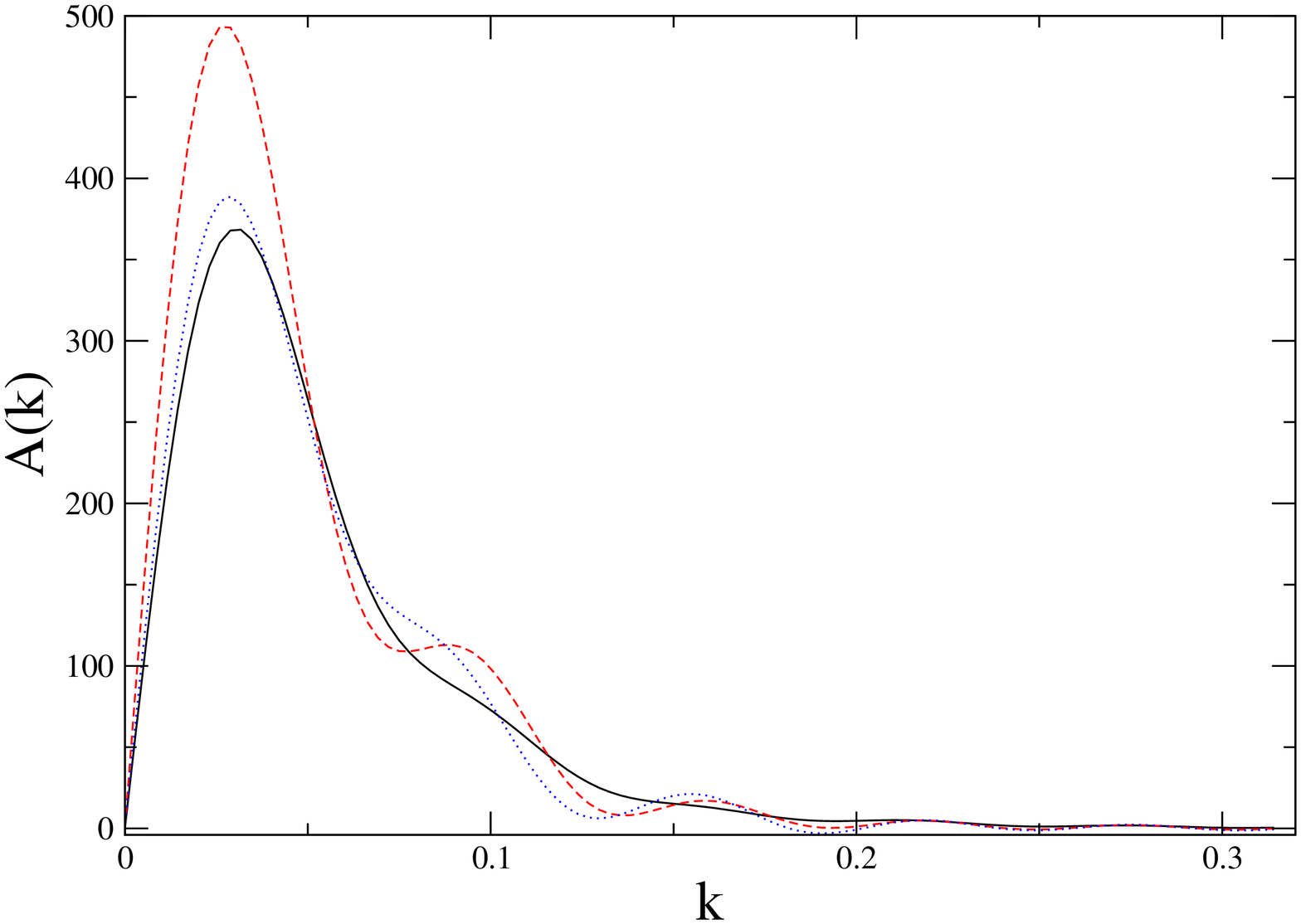}}
\caption{Plot of $A(k)$, analogous to Fig.~\ref{Sofk}. 
}
\label{Aofk}
\end{figure}

The Fourier space correlation functions can be found
using Eqs.~(\ref{SkfromMsfirst}), (\ref{AkfromMH}) and are
shown in Figs.~\ref{Sofk} and \ref{Aofk}. The spectra are
dominated by peaks at $k \sim 0.05$. This corresponds to a 
length scale $l \sim k^{-1} \sim 20$. From Fig.~\ref{nsphvst} 
we see that most sphalerons were nucleated at $t \sim 15$
and these would primarily be on bubbles that are also of 
size $\sim 15$, since most bubbles are nucleated at the 
earliest times. Hence the peak of the correlation is given 
by the sizes of the bubbles when bubbles started
to percolate. The width of the peaks in Figs.~\ref{Sofk} 
and \ref{Aofk} are $\Delta k \sim 0.05$ and also given
by the bubble sizes at percolation. Note that it would
not be suitable to characterize the injected spectra 
by power law fits.

\section{Conclusions and Future Directions}
\label{conclusions}

Following the general scenarios discussed in 
Refs.~\cite{Cornwall:1997ms,Vachaspati:2001nb} we have proposed 
a concrete model for the generation of helical magnetic fields 
during baryogenesis at a phase transition. The model takes into
account magnetic field generation due to baryon number violating 
processes occurring on bubble walls. By Monte Carlo simulations,
we have evaluated correlation functions of the injected magnetic 
field on completion of the phase transition. The Fourier space 
power spectra shown in Figs.~\ref{Sofk} and \ref{Aofk} tell us 
the characteristics of the magnetic fields injected into the 
plasma in this model. 

Our results should be viewed as providing initial conditions 
for subsequent evolution which will also entail MHD and cosmological 
effects. If the 
cosmological fluid is turbulent, say due to the motion of bubble 
walls, that too will play a role. These effects did not enter 
our study because the magnetic fields that
are produced due to sphaleron 
events are on scales comparable to the inverse W-boson mass
and far smaller than the scale at which the medium can be treated 
like a fluid. The micrscopic production, however, occurs at a high
rate, since it is also the rate at which baryons are produced, and 
the magnetic field due to different sphalerons will subsequently 
spread, merge and permeate space. 

MHD effects will come into play on length scales that are 
large compared to the thermal scale. As the magnetic field
expands to larger scales, but still less than the frozen-in 
scale, MHD effects will change the linear expansion to 
diffusive expansion. On yet larger scales, the magnetic field 
expansion enters the frozen-in regime where it can only scale 
with the expansion of the universe. In addition to these 
considerations, the evolution needs to include the helicity of 
the magnetic field and any turbulence that may accompany the
phase transition. It is known that helicity can be responsible
for an ``inverse cascade'' that transfers power to larger
scales. A discussion of some of these issues in the present
context may be found in Ref.~\cite{Vachaspati:2001nb}.

A long term goal of our model for generation of magnetic fields, 
is to connect the particle physics processes during baryogenesis
(such as the phase transition) to characteristics of the 
magnetic field. The hope is that eventually the observation
of a primordial magnetic field may say something about particle
physics at the baryogenesis scale, the nano-second universe,
and perhaps also the observed astrophysical magnetic structures.

\begin{acknowledgments}
This work was supported by the U.S. Department of Energy at 
Case Western Reserve University. TV was also supported by grant
number DE-FG02-90ER40542 at the Institute for Advanced Study.
\end{acknowledgments}

%\appendix


\begin{thebibliography}{43}

%\cite{Cornwall:1997ms}
\bibitem{Cornwall:1997ms}
  J.~M.~Cornwall,
  %``Speculations on primordial magnetic helicity,''
  Phys.\ Rev.\  D {\bf 56}, 6146 (1997)
  [arXiv:hep-th/9704022].
  %%CITATION = PHRVA,D56,6146;%%

%\cite{Vachaspati:2001nb}
\bibitem{Vachaspati:2001nb}
  T.~Vachaspati,
  %``Estimate of the primordial magnetic field helicity,''
  Phys.\ Rev.\ Lett.\  {\bf 87}, 251302 (2001)
  [arXiv:astro-ph/0101261].
  %%CITATION = PRLTA,87,251302;%%

%\cite{Manton:1983nd}
\bibitem{Manton:1983nd}
  N.~S.~Manton,
  %``Topology In The Weinberg-Salam Theory,''
  Phys.\ Rev.\  D {\bf 28}, 2019 (1983).
  %%CITATION = PHRVA,D28,2019;%%

%\cite{Klinkhamer:1984di}
\bibitem{Klinkhamer:1984di}
  F.~R.~Klinkhamer and N.~S.~Manton,
  %``A Saddle Point Solution In The Weinberg-Salam Theory,''
  Phys.\ Rev.\  D {\bf 30}, 2212 (1984).
  %%CITATION = PHRVA,D30,2212;%%

%\cite{Taubes:1982ie}
\bibitem{Taubes:1982ie}
  C.~H.~Taubes,
  %``The Existence Of A Nonminimal Solution To The SU(2) Yang-Mills Higgs
  %Equations On R**3,''
  Commun.\ Math.\ Phys.\  {\bf 86}, 257 (1982).
  %%CITATION = CMPHA,86,257;%%

%\cite{Vachaspati:1994ng}
\bibitem{Vachaspati:1994ng}
  T.~Vachaspati and G.~B.~Field,
  %``Electroweak string configurations with baryon number,''
  Phys.\ Rev.\ Lett.\  {\bf 73}, 373 (1994)
  [arXiv:hep-ph/9401220].
  %%CITATION = PRLTA,73,373;%%

%\cite{Hindmarsh:1993aw}
\bibitem{Hindmarsh:1993aw}
  M.~Hindmarsh and M.~James,
  %``The Origin of the sphaleron dipole moment,''
  Phys.\ Rev.\  D {\bf 49}, 6109 (1994)
  [arXiv:hep-ph/9307205].
  %%CITATION = PHRVA,D49,6109;%%

%\cite{Vachaspati:2006zz}
\bibitem{Vachaspati:2006zz}
  T.~Vachaspati,
  ``Kinks and domain walls: An introduction to classical and 
    quantum solitons,''
  {\it  Cambridge, UK: Univ. Pr. (2006) 176 p}

%\cite{Copi:2008he}
\bibitem{Copi:2008he}
  C.~J.~Copi, F.~Ferrer, T.~Vachaspati and A.~Achucarro,
  %``Helical Magnetic Fields from Sphaleron Decay and Baryogenesis,''
  Phys.\ Rev.\ Lett.\  {\bf 101}, 171302 (2008)
  [arXiv:0801.3653 [astro-ph]].
  %%CITATION = PRLTA,101,171302;%%

%\cite{Davidson:2008bu}
\bibitem{Davidson:2008bu}
  S.~Davidson, E.~Nardi and Y.~Nir,
  %``Leptogenesis,''
  Phys.\ Rept.\  {\bf 466}, 105 (2008)
  [arXiv:0802.2962 [hep-ph]].
  %%CITATION = PRPLC,466,105;%%

\bibitem{Jacksonbook}
 ``Classical Electrodynamics'',
 J.~D.~Jackson, Third Edition, Wiley (1998).

%\cite{Cline:2006ts}
\bibitem{Cline:2006ts}
  J.~M.~Cline,
  %``Baryogenesis,''
  arXiv:hep-ph/0609145.
  %%CITATION = HEP-PH/0609145;%%

%\cite{Bodeker:2009qy}
\bibitem{Bodeker:2009qy}
  D.~Bodeker and G.~D.~Moore,
  %``Can electroweak bubble walls run away?,''
  JCAP {\bf 0905}, 009 (2009)
  [arXiv:0903.4099 [hep-ph]].
  %%CITATION = JCAPA,0905,009;%%

%\cite{Espinosa:2010hh}
\bibitem{Espinosa:2010hh}
  J.~R.~Espinosa, T.~Konstandin, J.~M.~No and G.~Servant,
  %``Energy Budget of Cosmological First-order Phase Transitions,''
  arXiv:1004.4187 [hep-ph].
  %%CITATION = ARXIV:1004.4187;%%

%%\cite{Banerjee:2004df}
%\bibitem{Banerjee:2004df}
  %R.~Banerjee and K.~Jedamzik,
  %%``The Evolution of Cosmic Magnetic Fields: From the Very Early Universe, to
  %%Recombination, to the Present,''
  %Phys.\ Rev.\  D {\bf 70}, 123003 (2004)
  %[arXiv:astro-ph/0410032].
  %%%CITATION = PHRVA,D70,123003;%%

%\bibitem{Kolb90}
%E.~W.Kolb and M.~S.~Turner,
%\textit{The Early Universe.} Redwood City, CA: Addison-Wesley (1990)

%\bibitem{Arnold98}
%P.~Arnold, D.~Son and L.~Yaffe,
%Phys.\ Rev. {\bf D}{\bf 72}:054003 (2005)
%[arXiv:hep-ph/0505212] 

%%\cite{Hogan:1984hx}
%\bibitem{Hogan:1984hx}
  %C.~J.~Hogan,
  %%``Nucleation Of Cosmological Phase Transitions,''
  %Phys.\ Lett.\  B {\bf 133} (1983) 172.
  %%%CITATION = PHLTA,B133,172;%%

%%\cite{Baym:1995fk}
%\bibitem{Baym:1995fk}
  %G.~Baym, D.~Bodeker and L.~D.~McLerran,
  %%``Magnetic fields produced by phase transition bubbles in the electroweak
  %%phase transition,''
  %Phys.\ Rev.\  D {\bf 53}, 662 (1996)
  %%[arXiv:hep-ph/9507429].
  %%%CITATION = PHRVA,D53,662;%%

%%\cite{Neronov:2009gh}
%\bibitem{Neronov:2009gh}
  %A.~Neronov and D.~Semikoz,
  %%``Sensitivity of gamma-ray telescopes for detection of magnetic fields in
  %%intergalactic medium,''
  %arXiv:0910.1920 [astro-ph.CO].
  %%%CITATION = ARXIV:0910.1920;%%

\end{thebibliography}
\end{document}